\documentclass[doublecol]{epl2}
\usepackage{graphics}
\usepackage{color}

\begin{document}

\newcommand{\be}{\begin{equation}}
\newcommand{\ee}{\end{equation}}
\newcommand{\ra}{\rangle}
\newcommand{\la}{\langle}

\title{Semiclassical approach to universality in quantum chaotic transport}

\author{Marcel Novaes}

\institute{Departamento de F\'isica, Universidade Federal de S\~ao
Carlos, S\~ao Carlos, SP, 13565-905, Brazil}


\abstract{The statistics of quantum transport through chaotic
cavities with two leads is encoded in transport moments $M_m={\rm
Tr}[(t^\dag t)^m]$, where $t$ is the transmission matrix, which have
a known universal expression for systems without time-reversal
symmetry. We present a semiclassical derivation of this
universality, based on action correlations that exist between sets
of long scattering trajectories. Our semiclassical formula for $M_m$
holds for all values of $m$ and arbitrary number of open channels.
This is achieved by mapping the problem into two independent
combinatorial problems, one involving pairs of set partitions and
the other involving factorizations in the symmetric group.}

\maketitle

\section{Introduction}

A remarkable fact about electronic transport in ballistic systems
with chaotic classical dynamics is that they display universal
properties like, e.g., conductance fluctuations and weak
localization \cite{fluc}. These properties are well described by
random matrix theory (RMT) \cite{uzy,jala,rmt}, in which the system's details
are neglected and its $S$ matrix is modeled as a random element from
an appropriate matrix ensemble, specified only by the symmetries
present. It has always been a central problem to derive
such universal results from a semiclassical approximation in which
classical dynamics is taken into account. In this formulation,
quantum universality must emerge as a result of action correlations
that are present in any chaotic system: in essence, the existence of
long scattering trajectories that have nearly the same action and
interfere constructively in the semiclassical limit \cite{espalha}.
The same relation exists between long periodic orbits and spectral
statistics of closed systems \cite{berry}.

Consider a cavity with two leads attached, with $N_1$ and $N_2$ open
channels supported in each lead. This is taken as a model of actual
situations involving semiconductor quantum dots. Let $t$ be the
transmission block of the unitary $S$ matrix. The transport moments
$M_m={\rm Tr}[(t^\dag t)^m]$ carry much information about scattering
through the system. If time-reversal symmetry is broken by a strong
magnetic field, the RMT calculation of these quantities can be
carried out for arbitrary values of $N_1$ and $N_2$
\cite{prb78mn2008}. However, semiclassical derivations that agree
exactly with RMT predictions are available only for the first two
moments \cite{prl96sh2006,shot}, which are related to the
conductance and shot-noise of the cavity. Higher moments have been
treated perturbatively in $1/N$, where $N=N_1+N_2$, but only to
leading \cite{jpa41gb2008} and next-to-leading \cite{njp13gb2011}
orders.

In this article we remedy this situation, providing a semiclassical
derivation which is valid for all transport moments and to all
orders in perturbation theory, for broken time-reversal symmetry. Starting from a diagrammatic
formulation, we show how the diagrams that are relevant for $M_m$,
which involve scattering trajectories between leads, may be obtained
from closed diagrams involving only periodic orbits. Then, we map
them into certain factorizations of permutations in the symmetric
group. The factorizations required are different from the ones in
\cite{jpa41gb2008} (see also \cite{preprint}) and apparently have
not been considered before.

\section{Semiclassical Transport Moments}

In the semiclassical limit $\hbar\to 0$, $N\to\infty$, the matrix
elements of $t$ may be approximated \cite{c3hub1993} by \be
t_{oi}\approx\frac{1}{\sqrt{T_H}}\sum_{\gamma:i\to o}A_\gamma
e^{iS_\gamma/\hbar},\ee where the sum is over trajectories starting
at incoming channel $i$ and ending at outgoing channel $o$,
$S_\gamma$ is the action of trajectory $\gamma$ and $A_\gamma$ is an
amplitude related to its stability. The prefactor contains the
Heisenberg time $T_H$, which equals $N$ times the classical average dwell
time. Expanding the trace, transport moments become
\be\label{bigsum}
M_m\approx\frac{1}{T_H^m}\prod_{j=1}^m\sum_{i_j,o_j}\sum_{\gamma_j,\sigma_j}
A_{\gamma} A_{\sigma}^* e^{i(S_\gamma-S_\sigma)/\hbar}.\ee The sum
involves two sets of $m$ trajectories, the $\gamma$'s and the
$\sigma$'s. $A_\gamma=\prod_j A_{\gamma_j}$ is a collective
stability and $S_\gamma=\sum_j S_{\gamma_j}$ is a collective action,
and analogously for $\sigma$. Most importantly, the structure of the
trace implies that these two sets of trajectories connect the
channels in a different order, and we can arrange it so that
$\gamma_j$ goes from $i_j$ to $o_j$, while $\sigma_j$ goes from
$i_{j}$ to $o_{j+1}$.

The result of (\ref{bigsum}) is in general a strongly fluctuating
function of the energy, so a local energy average is introduced.
When this averaging is performed in the stationary phase
approximation, it selects those sets of $\sigma$'s that have almost
the same collective action as the $\gamma$'s. In the past 10 years
\cite{martin} it has been established that the mechanism behind
these action correlations is the existence of {\em encounters:} each
$\sigma$ must follow closely a certain $\gamma$ for a period of
time, and may exchange partners when some of the trajectories come together. A region where $\ell$ pieces of trajectories run
nearly parallel and $\ell$ partners are exchanged is called an
$\ell$-encounter. The two sets of trajectories are thus nearly
equal, differing only in the negligible encounter regions. In
particular, this implies $A_{\gamma}A_{\sigma}^*=|A_{\gamma}|^2$.
This theory has been presented with great detail in
\cite{haakepre,njp9sm2007}.

We show two examples of correlated sets of trajectories in Figure 1a,b, both contributing to $M_3$. Naturally, this is a simplification: in a realistic chaotic system these
trajectories would be long and extremely convoluted. The first example, Fig.1a, has a
triple encounter, within which the $\sigma$'s are represented by dashed lines. Before the encounter $\sigma_j$ is indistinguishable from $\gamma_j$; after the encounter it becomes indistinguishable from $\gamma_{j+1}$, i.e. the changing of partners inside the encounter
happens in such a way that $\sigma_j$ starts at $i_j$ and ends at $o_{j+1}$. Notice that
if the encounter was arranged differently, with $\sigma_j$ pairing up with
$\gamma_{j-1}$, this would not lead to an acceptable contribution to (\ref{bigsum}). The
second example, Fig.1b, has two double encounters. In this case each $\sigma$ ends up along the same $\gamma$ it started with. It is only acceptable as a
contribution to (\ref{bigsum}) when all outgoing channels coincide.

The present work will apply exclusively to systems for which the dynamics is not
invariant under time-reversal. This is because we shall not consider situations where a
$\sigma$ trajectory runs in the opposite sense with respect to a $\gamma$ trajectory.
This is a significant restriction on the possible correlated sets, and considerably
simplifies the treatment.

\begin{figure}[t]
\centerline{\includegraphics[scale=1.0,clip]{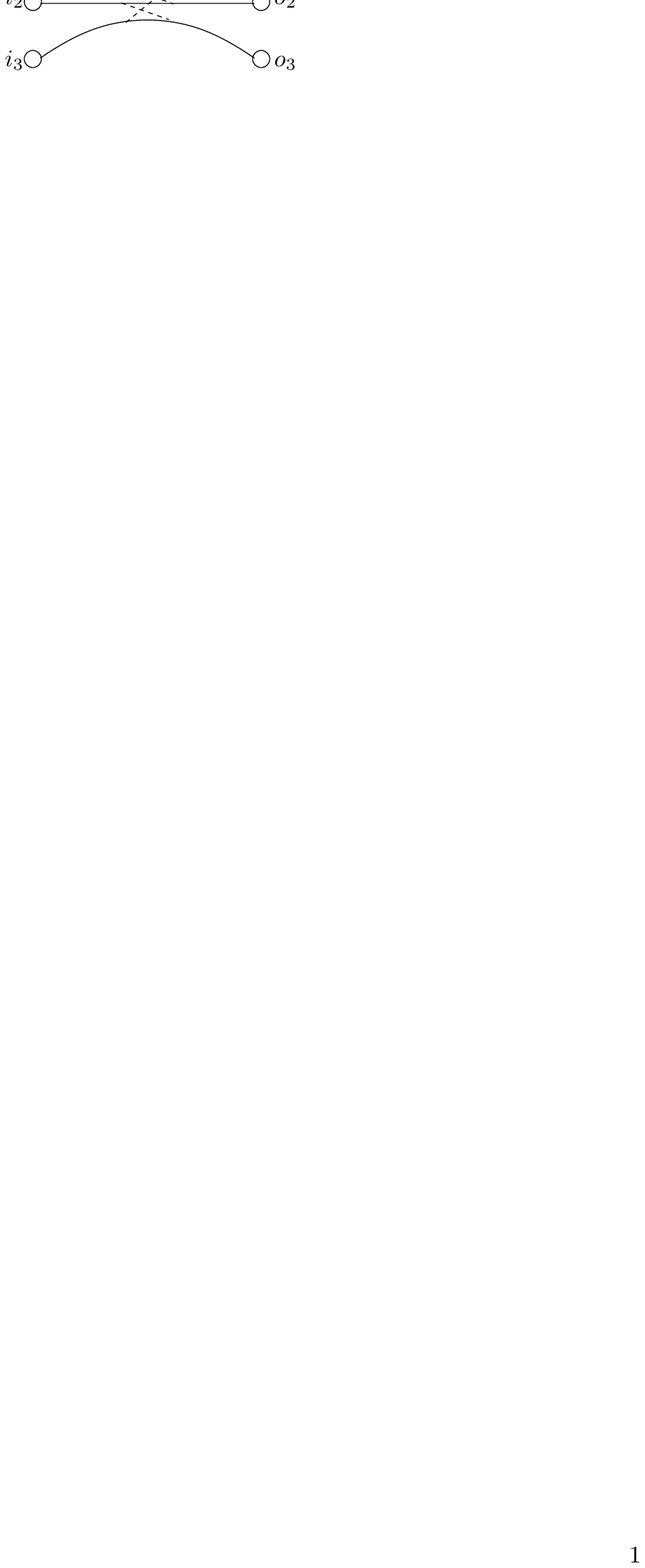}\hspace{0.2cm}\includegraphics[scale=1.0,clip]{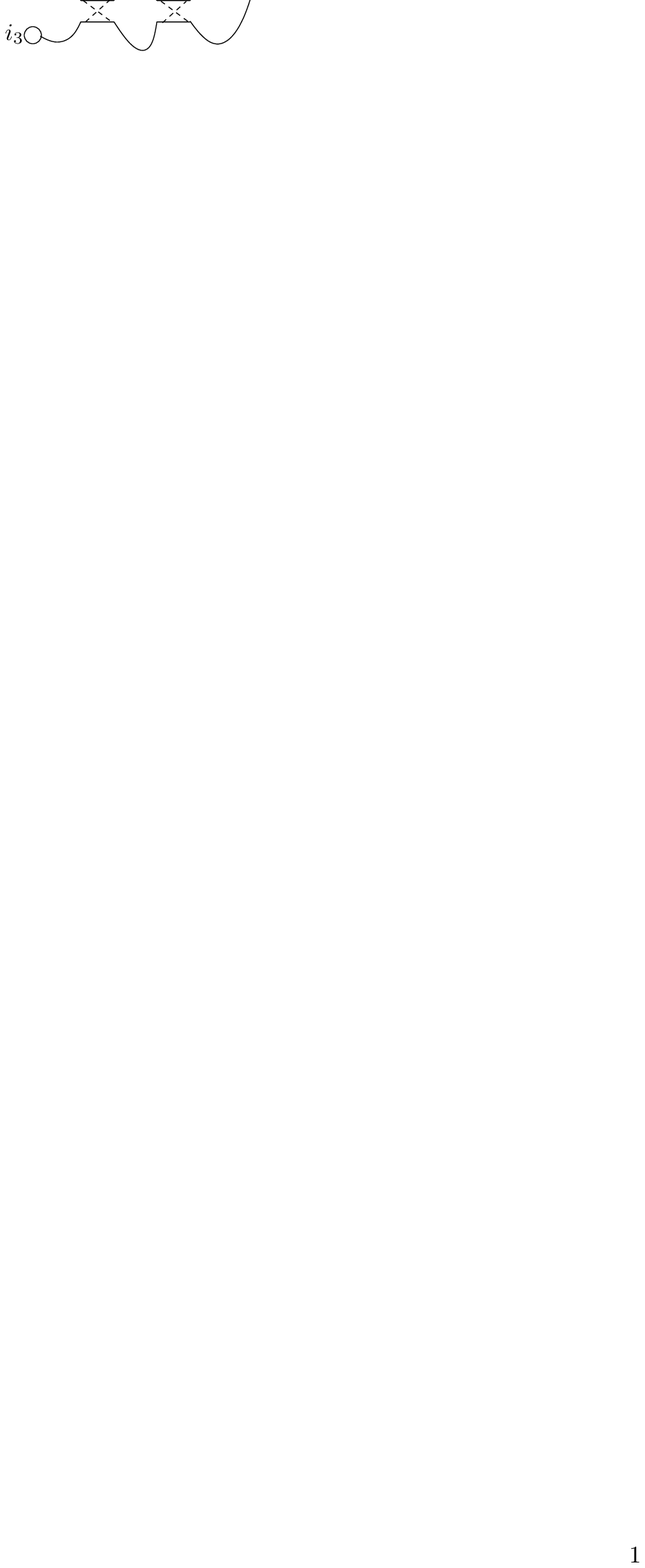}}
\centerline{\includegraphics[scale=1.1,clip]{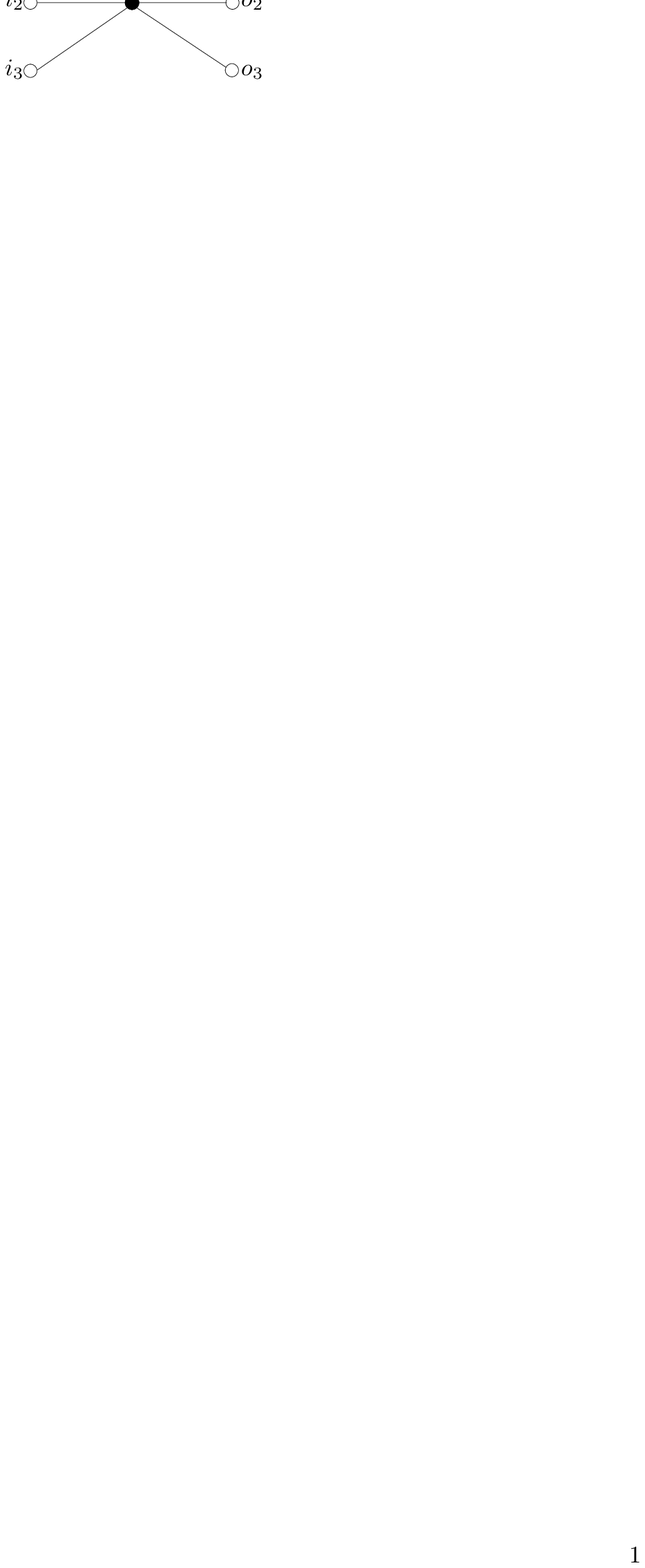}\hspace{0.4cm}\includegraphics[scale=1.1,clip]{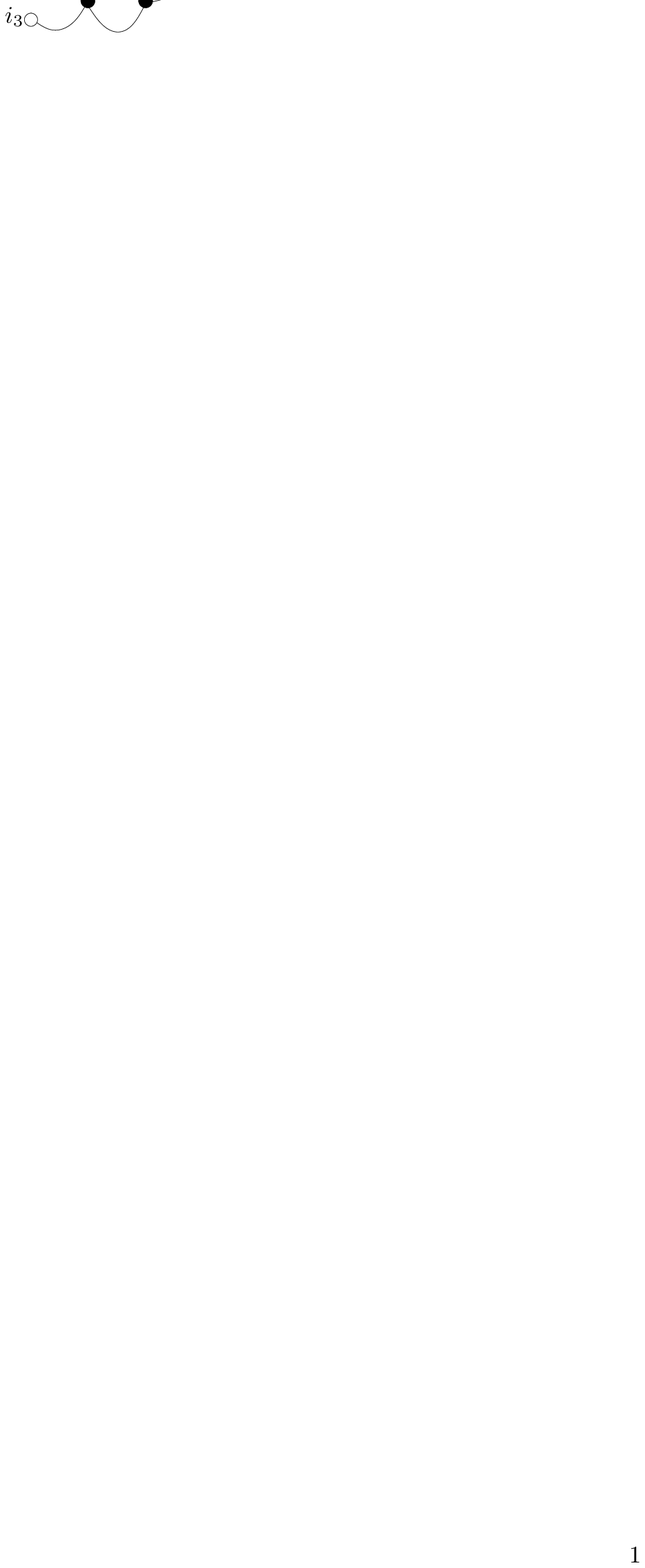}}
\caption{Two (pictorial) correlated sets contributing to $M_3$ are shown in a) and b).
The diagrams that correspond to them are shown in c) and d), respectively. The structure
of a) and c) is such that the $\sigma$ trajectories implement the permutation
$\pi=(123)$ on the channel labels. A different structure could be chosen so that $\pi=(132)$, but in that case c)
would be considered a pre-diagram, and some channels would have to coincide to make it a
diagram. Coinciding channels appear in b) and d), where only one structure is possible.}
\label{fig1}
\end{figure}

A correlated pair of trajectory sets may be represented by a {\it diagram} with
a {\it structure}. A diagram is a topological entity, a graph where encounters become vertices of
even valence and the pieces of trajectories leading from one encounter to another (or to
leads) become edges \cite{jpa41gb2008}. (Channels are formally also vertices, but they
should not be confused with encounters, so we will keep calling them channels.) The
structure is a prescription for walking in the graph, i.e. it specifies how the
$\sigma$'s change partners at the vertices. Examples are shown in Figure \ref{fig1}c,d.

It is known \cite{njp9sm2007} that the contribution of a diagram to transport moments
consists of $(-1)^VN^{V-E}\mathcal{C}(N_1,N_2)$, where $V$ is the number of vertices, $E$ is the number of edges and $\mathcal{C}(N_1,N_2)$ is the contribution due to the
channels. This factorizes into two parts. First, there is the number of ways they can be
assigned among the possible ones existing in the leads. If the number of {\em distinct}
incoming and outgoing channels are $m_1$ and $m_2$, then this is
\be\label{B}\mathcal{B}(m_1,m_2)=\frac{N_1!}{(N_1-m_1)!}\frac{N_2!}{(N_2-m_2)!}.\ee
Second, when there are $k$ different $\sigma$'s starting and ending at the same channels, there is a $k!$, since they could be labeled in that many different ways.

The diagram in Fig.\ref{fig1}a contributes as $-\mathcal{B}(3,3)/N^5$ to $M_3$. On the
other hand, the one in Fig.\ref{fig1}b contributes, also to $M_3$, as
$2\mathcal{B}(2,2)/N^5$, because $\sigma_2$ and $\sigma_3$ start and end at the same
channels, so there is freedom in choosing which is which.

It was shown in \cite{jpa41gb2008} that to leading order in $1/N$
all transport moments $M_m$ are determined by diagrams that have the
topology of a tree. This simplifying feature allowed a recursive
approach that lead to an explicit solution in agreement with RMT for
all $m$. Later, diagrams determining the first corrections were
constructed \cite{njp13gb2011} by grafting trees on base diagrams
with more complicated topology. We consider a different method,
inspired by the treatment of $M_1$ and $M_2$ in \cite{njp9sm2007}.
The crucial point is a relation between transport diagrams and
certain closed diagrams which have no channels but have one extra
vertex.

\section{Pre-diagrams}

\begin{figure}[t]
\center
\includegraphics[scale=1.1,clip]{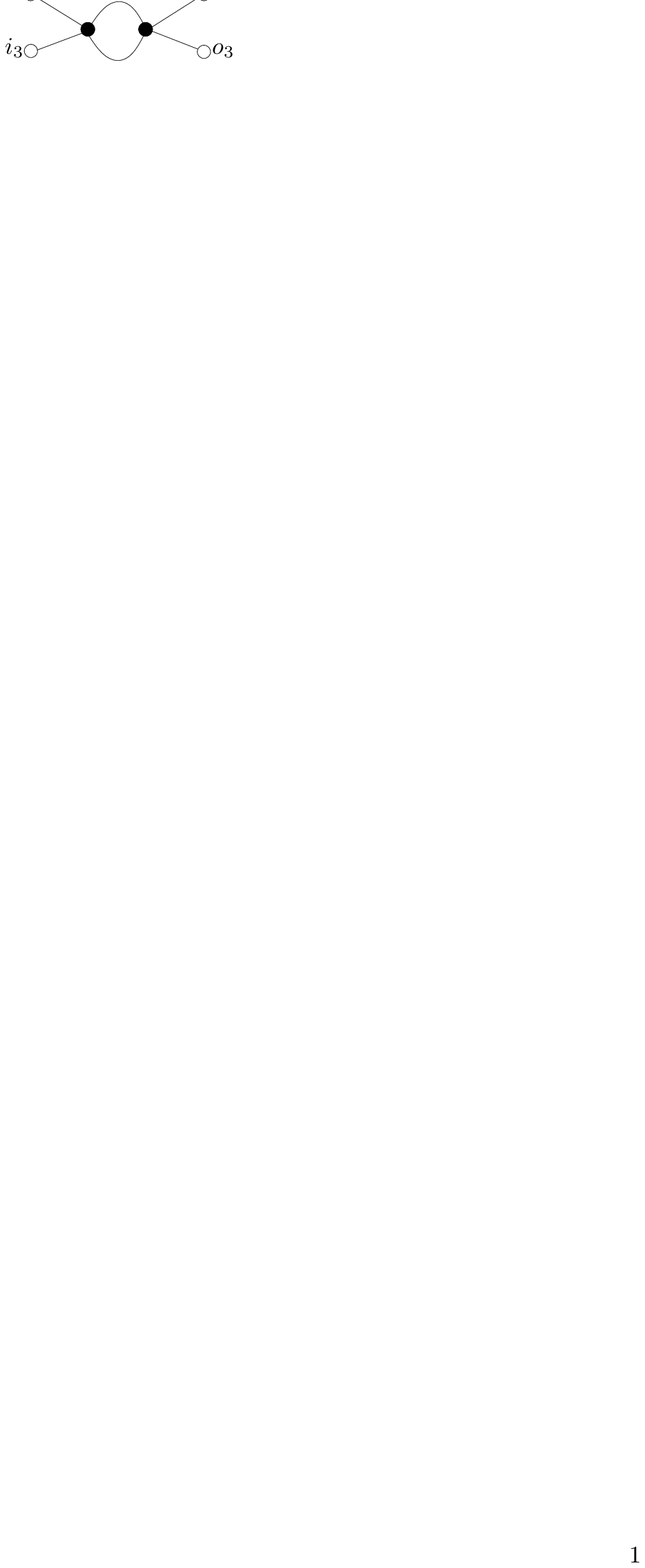}\hspace{1.0cm}
\includegraphics[scale=1,clip]{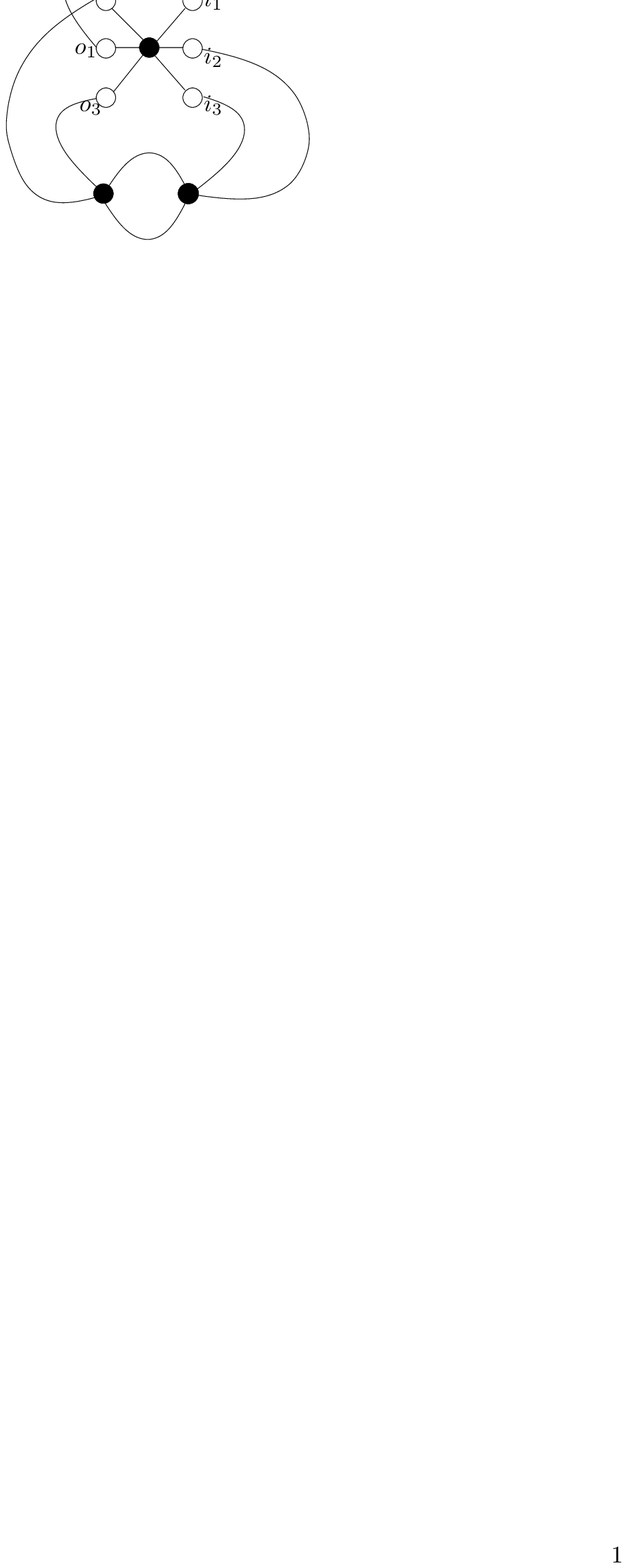}
\caption{ a) The pre-diagram corresponding to the diagram in
Fig.\ref{fig1}d. The permutation induced by the $\sigma$'s is the identity,
$\pi=(1)(2)(3)$. b) Introducing an extra vertex, the pre-diagram is
turned into correlated periodic orbits. A schematic representation
of the orbits is shown in Fig.\ref{fig4}.}
\label{fig2}
\end{figure}

We now describe our method, which draws on several combinatorial
ideas. Details will be presented in a future publication. Let $S_m$
denote the symmetric group of order $m$, i.e. the set of all
permutations of $m$ symbols. We denote by $1_m$ the identity in
$S_m$ and by $c_m=(12\cdots m)$ the complete cyclic
permutation with increasing elements. Note that $c_m$ is the permutation the $\sigma$'s are
required to perform on the channel labels, i.e. they should take
$i_j$ to $o_{j+1}$. On the other hand, the $\gamma$'s perform $1_m$ by taking $i_j$ to $o_j$.

Let us introduce a class of graphs we call {\em pre-diagrams}, which have $m$ incoming
and $m$ outgoing channels, assumed all distinct. We call $m$ the order of the
pre-diagram. Trajectory $\gamma_j$ is still required to go from $i_j$ to $o_j$, but
$\sigma_j$ is allowed to go from $i_{j}$ to {\em any} outgoing channel. This means that
in a pre-diagram the permutation performed by the $\sigma$'s on the channel labels is
generally different from $c_m$. Let $\pi$ denote this permutation. We present an example
in Figure \ref{fig2}a. Clearly, to every given diagram with given structure we can
associate a unique pre-diagram (obtained simply by ignoring coincidences between
channels).

There are therefore two important differences between a diagram and its pre-diagram.
First, in a diagram trajectory $\sigma_j$ start at $i_j$ and end at $o_{j+1}$,
while in a pre-diagram it can end anywhere and the set $\sigma$ may implement any permutation $\pi$ on the channel labels. Second, there may be coinciding
channels in a diagram, while they are all different in a pre-diagram. The idea is that
true diagrams can be obtained from a pre-diagram by means of those coincidences among
channels that make $\pi$ effectively equivalent to $c_m$. For example, if $\pi=1_m$ then
several coincidences are needed, such as all incoming or all outgoing channels. In order
to count all diagrams, we must count all possible coincidences for all possible
pre-diagrams.

\begin{figure}[t]
\centerline{\includegraphics[scale=1,clip]{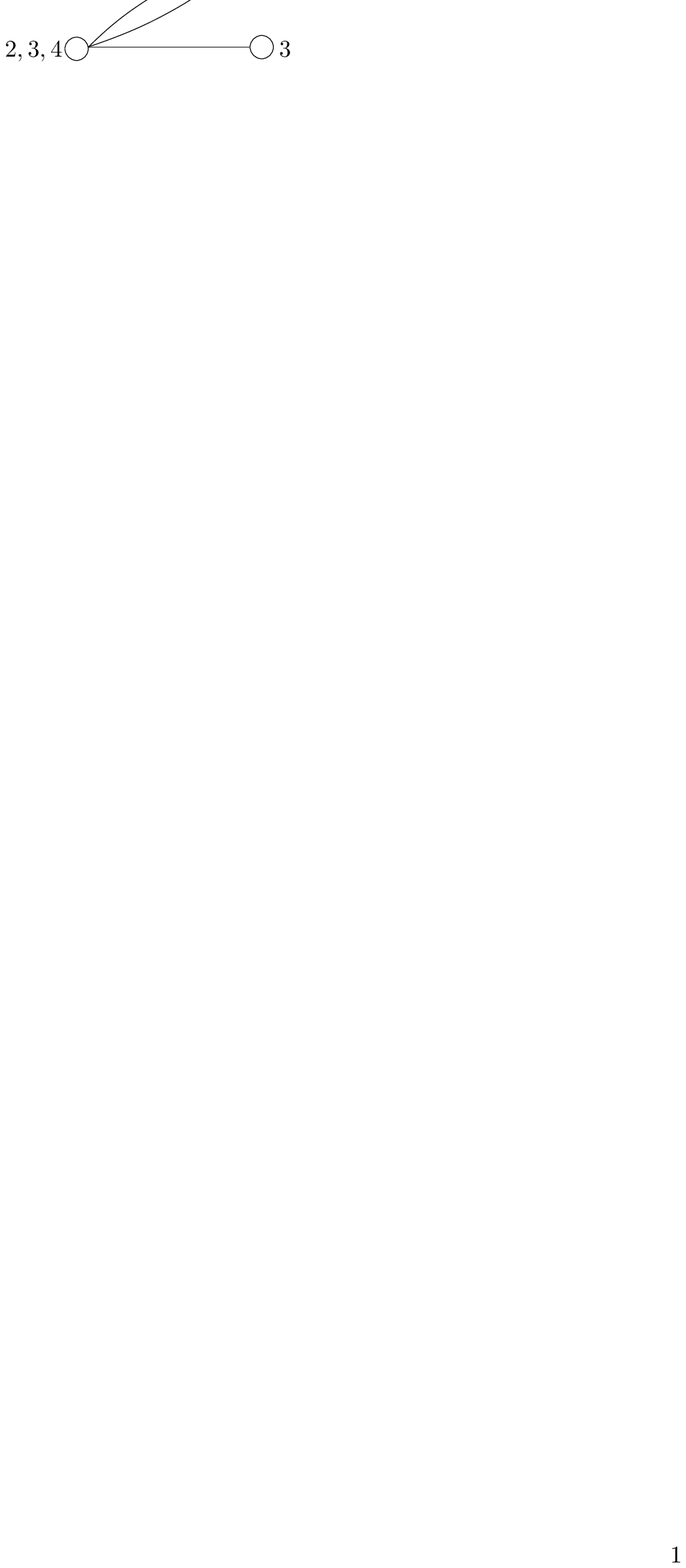}} \caption{The
partitions in this graph constitute a coincidence pair admissible under the
permutation $\pi=(12)(34)$. This is because the bipartite graphs
induced by connecting the blocks according to
$\pi$ or according to $(1234)$ are the same. The multiplicity of
this graph is $2$, because the trajectories represented by the
double bond admit two possible labelings.}
\label{fig3}
\end{figure}

\section{Admissible pairs of partitions}

Let us formalize the above ideas. Coinciding channels induce partitions of the set $\{1,\dots,m\}$ on the incoming and outgoing leads. Suppose two sets of $m$ points arranged
vertically side by side. Given a partition $L$ of the left set and a partition $R$ of the
right set, identify the points according to the blocks of the partitions. Given a
permutation $\pi\in S_m$, draw lines going from the left points to their images under
$\pi$ on the right. If $L$ and $R$ have $m_1$ and $m_2$ blocks respectively, this
produces a bipartite graph $\mathcal{G}(L,R;\pi)$ with $m_1$ vertices on the left and
$m_2$ vertices on the right. An example is shown in Figure \ref{fig3}, where the left
partition is $L=\{\{1\},\{2,3,4\}\}$, the right one is $R=\{\{1,2,4\},\{3\}\}$ and the
permutation is $\pi=(12)(34)$.

We say that $L$ and $R$ form a $\pi$-admissible coincidence pair if
$\mathcal{G}(L,R;\pi)=\mathcal{G}(L,R;c_m)$. Physically, this means that the permutation
induced by the $\sigma$'s is `effectively' equal to $c_m$, and the diagram contributes to
(\ref{bigsum}). Let $\mathcal{M}(L,R;\pi)$ be the incidence matrix of
$\mathcal{G}(L,R;\pi)$. This means that $\mathcal{M}_{jk}$ is the number of bonds going
from block $j$ on the left to block $k$ on the right. The incidence matrix of the graph
in Fig.\ref{fig3} is $\left(\begin{array}{cc}1 & 0 \\2 & 1 \end{array}\right)$. As we
have discussed, each multiple bond gives rise to a factorial because the $\sigma$
trajectories could be interchanged. We thus define the multiplicity of the pair $(L,R)$
as \be \mu(L,R;\pi)=\prod_{jk}[\mathcal{M}(L,R;\pi)]_{jk}!\ee

Let $\mathcal{A}(\pi,m_1,m_2)$ be the set of all $\pi$-admissible coincidence pairs with
$m_1$ blocks on the left and $m_2$ blocks on the right. The function \be
f(\pi,m_1,m_2)=\sum_{(L,R)\in\mathcal{A}(\pi,m_1,m_2)}\mu(L,R;\pi)\ee counts, with the
correct multiplicity, diagrams with $m_1$ distinct channels on the left and $m_2$
distinct channels on the right, which have pre-diagram with permutation $\pi$. Once it is
known, we can compute \be
F(\pi,N_1,N_2)=\sum_{m_1,m_2}f(\pi,m_1,m_2)\mathcal{B}(m_1,m_2),\ee where
$\mathcal{B}(m_1,m_2)$ is given by (\ref{B}). The quantity above is the total
contribution to the sum (\ref{bigsum}) associated with pre-diagrams for which the
permutation performed by the $\sigma$'s on channel labels is $\pi$. Notice that since
$\pi\in S_m$ this depends implicitly on $m$.

We therefore meet our first combinatorial problem, which is to obtain the function
$F(\pi,N_1,N_2)$, or at least make it computable via a generating function or a
recurrence relation. This requires understanding the set $\mathcal{A}(\pi,m_1,m_2)$,
which seems to be a very complicated combinatorial problem. Once we have $F(\pi,N_1,N_2)$
under control, we are left with the problem of finding all possible pre-diagrams.

\section{Relation to correlated periodic orbits}

\begin{figure}[t]
\centerline{\includegraphics[scale=0.9,clip]{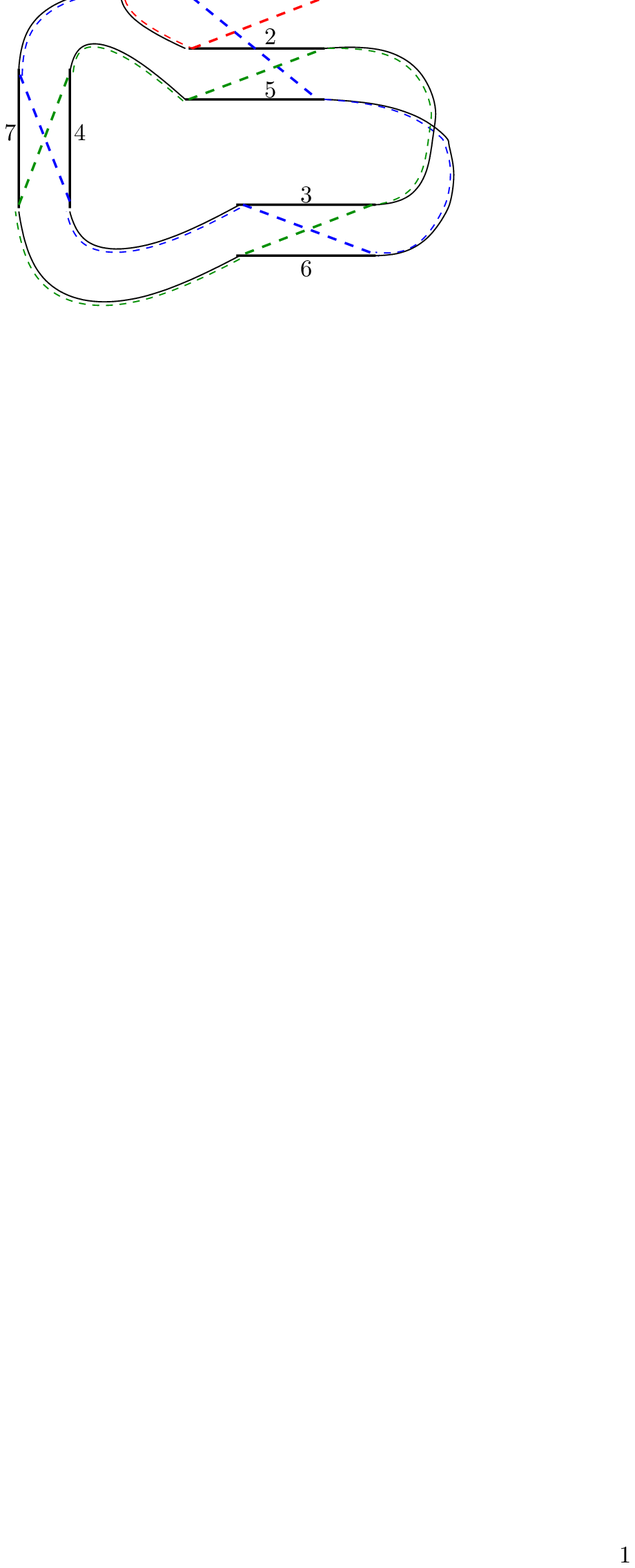}}
\caption{(color online) Schematic representation of the correlated
periodic orbits in Fig. 2(b), one depicted with solid line and three
with dashed lines. With this choice of labels the associated
factorization equation in the symmetric group $S_7$ is
$(1234567)=(1)(264)(375)\cdot(125)(36)(47)$. Removing the
$3$-encounter we produce the pre-diagram in Fig.2(a), such that trajectory $\gamma_1$ has no stretches while trajectories $\gamma_2$ and $\gamma_3$ include stretches \{3,4\} and \{5,6\}, respectively.}
\label{fig4}
\end{figure}

We now turn to the question of counting pre-diagrams. First, we
associate pre-diagrams with certain collections of correlated
periodic orbits. Let $\alpha$ be a single periodic orbit and let
$\beta$ be a set of periodic orbits that is correlated with
$\alpha$, i.e. $\beta$ differs from $\alpha$ only in encounters.
We show an example in Figure \ref{fig4}, where one periodic orbit is correlated with three others, in a situation with two $2$-encounters and one $3$-encounter.

Given $\alpha$ and $\beta$, suppose we `cut open' an $m$-encounter. This produces $2m$ endpoints, $m$ of them corresponding to the `beginning' of trajectories (leaving the encounter) and another $m$ to `ending' of trajectories (arriving at the encounter). We interpret them as incoming and outgoing channels, respectively. Then, we choose one of
the incoming channels to be $i_1$, and use $\alpha$ to label all
channels in sequence: the piece of $\alpha$ that starts in $i_j$
(and necessarily ends in $o_j$) becomes $\gamma_j$, while the
piece of $\beta$ that starts in $i_j$ becomes $\sigma_j$. This
obviously produces a pre-diagram, with a certain permutation $\pi$
to be determined. See Figure 2b.

There are two caveats in the procedure outlined above. The first is
that some of the $\beta$'s might not participate in the encounter we
chose to cut open, and would not become scattering trajectories. This is
remedied simply by demanding that in our correlated pair $\alpha,\beta$ there must an $m$-encounter in which all $\beta$'s participate. Second, even though any pre-diagram
related to $M_m$ can be produced in this way, some of them could be
produced more than once. We show the simplest possible such example in Figure 5: the same pre-diagram could be produced having three $\beta$'s or only one $\beta$. However, the structure of the encounter that was opened is necessarily different. In order to avoid this overcounting, we have at our disposal the possibility to select the permutation experienced by the $\sigma$'s inside the encounter that was opened. We shall make
use of this right after Eq.(\ref{xi}).

In \cite{haakepre}, pairs of correlated periodic orbits were associated with
factorizations of permutations. We summarize the idea. Suppose some correlated pair
$\alpha$, $\beta$. Label the encounter stretches in such a way that the end of stretch $j$ is followed by the beginning of stretch $j+1$. This produces the permutation $c_E$, where $E$ is the number of stretches, acting on the `exit-to-entrance' space (it goes from the exit of an encounter to the entrance of another one). A variant of this construction is shown in Fig.\ref{fig4}, where $\alpha$ is represented by a solid line.

The orbits behave differently at the encounters (the `entrance-to-exit'
space). At any encounter $\alpha$ corresponds to the identity permutation, since it takes the entrance of a stretch to the exit of the same stretch. On the other hand, $\beta$ may be represented by a non-trivial permutation $P$, whose number of cycles is equal to $V$, the number of encounters. In the example shown in Figure 4 we have $P=(125)(36)(47)$. The product $c_E P=Q$, acts on `exit-to-exit' space, leading from the exit of an encounter to the exit of another one. It must be a single cycle if there is a single $\beta$. Since $c_E$ is fixed, the total number of correlated pairs equals the number of solutions in $S_E$ to the factorization $c_E=QP^{-1}$.

In our problem we have a single periodic orbit $\alpha$ which is correlated to a set of
any number of periodic orbits, $\beta$. It is possible to adapt this mapping into
permutations to aid us in the enumeration of pre-diagrams. We have to remove the
condition that the permutation $Q$ be single-cycle, to allow for more than one $\beta$.
In the example shown in Figure 4 there are three different $\beta$'s, and we have $Q=(1)(264)(375)$. We must also ensure that all $\beta$'s take part in the encounter we open, which we convention to be the first one. Finally, we have to make sure that when we produce the pre-diagram, the permutation implemented by $\sigma$ on the channel labels is given by $\pi$. When we remove the encounter involving stretches $\{1,2,5\}$ in Figure 4, we produce a pre-diagram for which $\pi$ is the identity in $S_3$.

\section{Some factorizations of permutations}

Let $\{P\}$ denote the set of integers which
are not fixed points of the permutation $P$. Let $P_1$ denote the first cycle of $P$, the
one that contains the element `$1$'. Given a set $s$, we define $\left.P\right|_s$, the
restriction of $P$ to $s$, to be the permutation obtained by simply erasing from the
cycle representation of $P$ all symbols not in $s$. For example,
$\left.(123)\right|_{\{1,3\}}=(13)$.

We also define a slightly more involved operation we call reduction. Given a permutation
$Q$ and a set $s$, the reduction $\mathcal{R}_s[Q]$ is obtained by first restricting $Q$
to $s$ and then making each element as small as possible keeping positivity and relative
order. For example, $\mathcal{R}_{\{2,3,4,6\}}[(264)]=(143)$. This is found as follows.
First, because we are restricting to $\{2,3,4,6\}$, we must write explicitly the fixed
point: $(264)(3)$. Then reduction leads to $(143)(2)$. Finally, we may omit again the
fixed point. Another example: $\mathcal{R}_{\{2,3,4\}}[(264)]=(13)$.

Suppose we have correlated orbits described by the equation $c_E P=Q$. The set of elements involved in the first encounter is $\{P_1\}$, assumed to have $m$ elements. In the example of Fig.\ref{fig4} this is $\{1,2,5\}$. The $\gamma$ trajectories start and end at this encounter and, by construction, visit these elements in increasing order, i.e. they implement a permutation which is simply $\left.c_E\right|_{\{P_1\}}$. This is $(125)$ in Fig.\ref{fig4}.

We must determine what is $\pi$, the permutation induced by $\sigma$ on those labels. First, we take account the permutation $Q$ and reduce it to the appropriate space, $\left.Q\right|_{\{P_1\}}$. In Fig.\ref{fig4} this is $(1)(2)(5)$. This acts on exit-to-exit space, i.e. it takes incoming channels to incoming channels. We must therefore multiply it by the inverse of $P_1$ in order to reverse the permutation effected inside the first encounter. The result, $\left.Q\right|_{\{P_1\}}P_1^{-1}$, takes incoming channels to outgoing channels. In Fig.\ref{fig4} this is also $(125)$, just like for the $\gamma$'s.

\begin{figure}[t]
\centerline{\includegraphics[scale=0.4,clip]{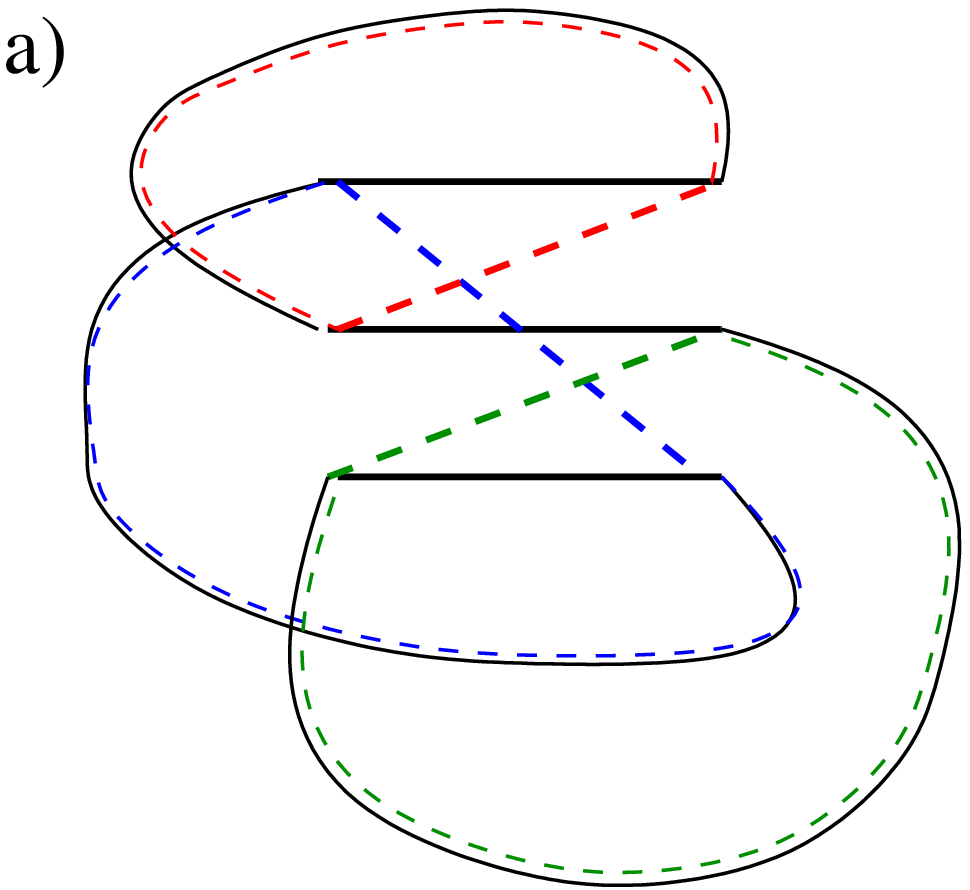}\hspace{0.5cm}\includegraphics[scale=0.4,clip]{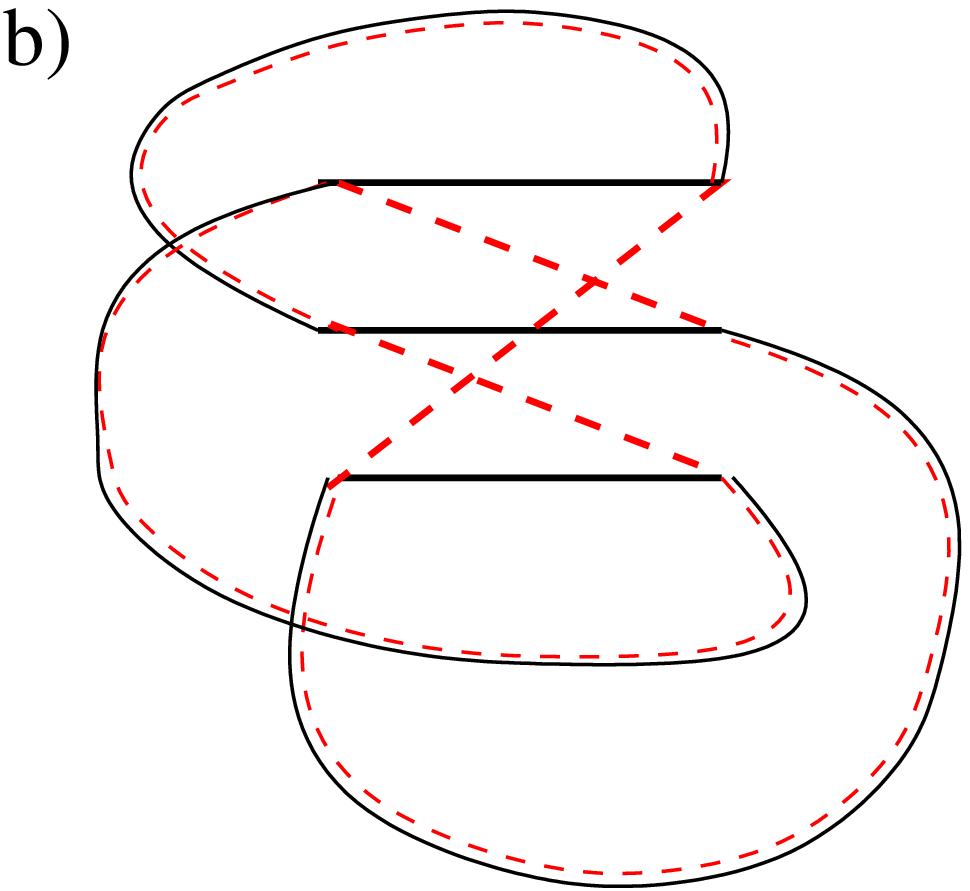}}
\caption{(color online) a) Schematic representation of a periodic orbit (solid line)
which is correlated with three periodic orbits (dashed lines). Removing the encounter we
produce a pre-diagram with permutation $\pi=(1)(2)(3)$. Notice that the permutation
inside the encounter, $(132)\equiv (321)$, is decreasing. b) Schematic representation of
a pair of correlated periodic orbits. Removing the encounter would produce the same
pre-diagram as a), but we discard this situation because the permutation inside the
encounter, $(123)$, is not decreasing.} \label{fig5}
\end{figure}

At this point, we have the permutations implemented by both $\gamma$ and $\sigma$ on the channel labels. This first is $\left.c_E\right|_{\{P_1\}}$ and the second is $\left.Q\right|_{\{P_1\}}P_1^{-1}$. The permutation $\pi$ corresponds to measuring the second one {\em with respect} to the first one. In other words, we must carry out a change of coordinates of sorts. We therefore multiply both quantities by the inverse of
the first. This turns the second one into $\left.Q\right|_{\{P_1\}}
P_1^{-1}\left.c_E^{-1}\right|_{\{P_1\}}$. In our example of Fig.\ref{fig4}, this gives $(1)(2)(5)$. The permutation
$\pi\in S_m$ is finally obtained after reduction: \be\label{xi}
\pi=\mathcal{R}_{\{P_1\}}\left[\left.Q\right|_{\{P_1\}}
P_1^{-1}\left.c_E^{-1}\right|_{\{P_1\}}\right].\ee

As we have mentioned, some pre-diagrams can be produced more than once. A simple example
is shown in Figure \ref{fig5}. In order to avoid this overcounting we may impose a convention on the permutation experienced by the $\sigma$'s inside the encounter which is cut open. A very convenient choice is to demand that $P_1^{-1}=\left.c_E\right|_{\{P_1\}}$,
which means that $P_1^{-1}$ is increasing (i.e. its elements are ordered increasingly)
or, equivalently, that $P_1$ is decreasing (i.e. its elements are ordered decreasingly,
as in Fig.\ref{fig5}a). This leads to a much simpler expression for the permutation $\pi$: \be \pi=\mathcal{R}_{\{P_1\}}\left[Q\right].\ee In particular, the number of cycles of $\pi$ equals the number of individual periodic orbits in the set
$\beta$.

All in all, the combinatorial problem that needs to be solved is the following. One must
find the number of solutions in $S_E$, let us denote it by $\Xi(m,\pi,E,V)$, to the
factorization equation $c_E=QP^{-1}$ which satisfy several conditions: i) $P$ has $V+1$
cycles; ii) $P_1^{-1}$ is increasing and of size $m$; iii) all cycles of $Q$ have at
least one element in common with $P_1$; iv) the reduction of $Q$ to $\{P_1\}$ is equal to a given $\pi$.

We therefore meet our second combinatorial problem, which is to obtain the function
$\Xi(m,\pi,E,V)$, or at least make it computable via a generating function or a
recurrence relation. As we have seen, this requires handling a very involved
factorization problem in the symmetric group.

\section{Conclusions}

We have shown that the semiclassical calculation for transport moments results in
\be\label{final} M_m=\sum_{\pi\in S_m} F(\pi,N_1,N_2)\sum_{E,V}
\Xi(m,\pi,E,V)\frac{(-1)^V}{N^{E-V}},\ee where the quantities involved have been defined
in the text. This expression is exact, i.e. valid for arbitrary numbers of channels. It
is easy to implement it in symbolic packages and verify that it indeed reproduces RMT
results, as far as it can be checked. Unfortunately, at present both combinatorial
problems presented here remain open, so that a closed form expression for $M_m$ is beyond
reach. However, we stress that the semiclassical approach is not intended as a
computational tool. Its merit is in revealing the dynamical origin of universality.

Exact semiclassical results are somewhat surprising in the regime of
finite number of channels, specially for high moments, because the
number of channels in a lead is small when its width is of the order
of the electronic wavelength, when diffraction effects are expected
to be important \cite{rotter}. It is not clear what role is played
by diffraction in bringing about universal statistics in chaotic
geometries, and how it would affect the semiclassical description
presented here. Another point that remains to be addressed are the
corrections arising when the system's Ehrenfest time $T_E$ is larger
than the typical dwell time $T_D$ in the cavity (we have assumed
$T_E\ll T_D$, which is when RMT universality is expected). Some
progress has been made in that direction \cite{TE}, but only to
leading order in $1/N$. These issues deserve further study.

In summary, we have presented a semiclassical derivation of the universal statistics of
quantum transport in chaotic systems. This was achieved by reducing the problem to two
independent combinatorial questions. One of them involves certain pairs of set partitions
and is embodied in the function $F(\pi,N_1,N_2)$. The other involves certain
factorizations in $S_E$ and leads to the function $\Xi(m,\pi,E,V)$. We hope that the
present article will draw attention to the profound relationship between quantum chaos
and combinatorics. Similar ideas can be developed for closed systems, to provide all
spectral correlation functions. Work in this direction is underway \cite{closed}.

\acknowledgments

This work was supported by FAPESP and CNPq. After it was completed I learned that
Berkolaiko and Kuipers also found a semiclassical derivation of $M_m$ valid for all $m$
and all channel numbers \cite{also}.


\end{document}